\documentclass[twocolumn]{aastex631}

\usepackage{hyperref}

\newcommand{\ha}{${\rm H\alpha}$}

\newcommand{\pa}{${\rm Pa\alpha}$}
\newcommand{\nii}{\hbox{[N\,{\sc ii}]}}
\usepackage{threeparttable}

\begin{document}

\title{Deciphering Gas Dynamics and Star Formation in a z=1.1 Main Sequence Spiral Galaxy with ALMA and JWST}

\correspondingauthor{Zhaoran Liu}
\email{zhaoran.liu@astr.tohoku.ac.jp}
\author[0009-0002-8965-1303]{Zhaoran Liu}
\affiliation{Astronomical Institute, Graduate School of Science, Tohoku University, 6–3 Aoba, Sendai 980-8578, Japan}

\author[0000-0002-2993-1576]{Tadayuki Kodama}
\affiliation{Astronomical Institute, Graduate School of Science, Tohoku University, 6–3 Aoba, Sendai 980-8578, Japan}

\author[0000-0002-8512-1404]{Takahiro Morishita}
\affiliation{IPAC, California Institute of Technology, MC 314-6, 1200 E. California Boulevard, Pasadena, CA 91125, USA}

\author[0000-0003-4814-0101]{Kianhong Lee}
\affiliation{Astronomical Institute, Graduate School of Science, Tohoku University, 6–3 Aoba, Sendai 980-8578, Japan}
\affiliation{National Astronomical Observatory of Japan, 2-21-1 Osawa, Mitaka, Tokyo 181-8588, Japan}

\author[0000-0002-4622-6617]{Fengwu Sun}
\affiliation{Center for Astrophysics $|$ Harvard \& Smithsonian, 60 Garden St., Cambridge, MA 02138, USA}

\author[0000-0002-7598-5292]{Mariko Kubo}
\affiliation{Astronomical Institute, Graduate School of Science, Tohoku University, 6–3 Aoba, Sendai 980-8578, Japan}

\author[0000-0001-8467-6478]{Zheng Cai}
\affiliation{Department of Astronomy, Tsinghua University, Beijing 100084, China}

\author[0000-0003-0111-8249]{Yunjing Wu}
\affiliation{Department of Astronomy, Tsinghua University, Beijing 100084, China}

\author[0000-0001-5951-459X]{Zihao Li}
\affiliation{Cosmic Dawn Center (DAWN), Denmark}
\affiliation{Niels Bohr Institute, University of Copenhagen, Jagtvej 128, DK-2200, Copenhagen N, Denmark}

\begin{abstract}
We present a joint analysis of high-resolution CO(2-1) and Paschen-$\alpha$ (Pa$\alpha$) emission lines to trace gas dynamics and spatially resolved star formation in ASPECS-LP.3mm.06, a $z=1.1$ main sequence galaxy. Utilizing data from the ALMA and JWST NIRCam Wide Field Slitless Spectroscopy (WFSS), we explore both ionized gas and molecular gas within this galaxy. With a substantial molecular gas fraction (f$_\mathrm{mol}$ = 0.44 $\pm$ 0.02), ASPECS-LP.3mm.06 remains on the star-forming main sequence and adheres to the Kennicutt-Schmidt (KS) relation, indicating typical gas-to-star conversion efficiency. Our analysis reveals extended structures across multiple wavelengths, suggesting regulated star formation within a stable disk. The spatially resolved star formation efficiency (SFE) and kinematic analysis indicate that ASPECS-LP.3mm.06 features a smooth mass assembly process across bulge and disk. Additionally, the galaxy exhibits modest dust extinction (A$_\mathrm{V}$ = 0.8), potentially linked to self-regulation during bulge formation. These findings position ASPECS-LP.3mm.06 as a prototypical galaxy, offering valuable insights into the mechanisms governing normal disk galaxy growth at z$\sim$1.
\end{abstract}



\section{Introduction} \label{sec:intro}
Galaxies form and evolve through the intricate dynamics of dark matter halos and baryonic matter. Dark matter halos serve as the gravitational backbone, attracting and stabilizing baryonic matter, thus facilitating galaxy formation. Within these halos, gas cools and condenses, forming stars that shape the varied and complex structures of galaxies. At the heart of this process is the transformation of gas into stars. Molecular gas accumulates in cold, dense regions known as molecular clouds. These clouds, often referred to as stellar nurseries, are where gas cools and condenses under the influence of gravity. As the gas continues to collapse, it eventually ignites nuclear fusion, giving birth to new stars. As these newborn stars emerge, particularly the massive ones, they begin to influence their surroundings in profound ways. The intense ultraviolet (UV) radiation from these young stars ionizes the nearby hydrogen gas, creating luminous H{\footnotesize II} regions. These pockets of ionized gas illuminate the story of recent star formation, marking the locations where stars have just been born \citep{Tinsley80, Mo98, Carilli13, Tacconi20}. 
Dust is a vital component in these star-forming regions, it plays a crucial role within molecular clouds by shielding gas from UV radiation, which allows the gas to cool and form molecules, ensuring that these clouds remain cold and dense enough to support star formation \citep{Calzetti2000, Dwek11}. In H{\footnotesize II} regions, dust grains absorb and scatter the UV light from young stars. This interaction is essential for cooling the gas, as dust re-emits the absorbed energy in the infrared spectrum. Thus, the amount of dust is directly linked to star formation activities, e.g., merger-driven starburst galaxies are expected to exhibit higher dust attenuation in their centers due to the concentration of star-forming regions \citep{Bekki2000}, while galaxies experiencing smooth gas inflow-driven star formation tend to have more uniformly distributed dust and less central attenuation. 

With its high spatial and spectral resolution capabilities, ALMA has significantly advanced our understanding of molecular gas and dust in galaxies. By probing both the dust continuum and gas emission lines, ALMA offers detailed insights into the distribution of dust and star formation processes across diverse galactic environments and redshifts \citep[e.g.,][]{Simpson15, Hodge16, Tadaki17, Elbaz18, Calistro18, Hodge19,Morishita22, Wu23, Lee24}. These observations have allowed us to visualize critical components and understand their essential roles in regulating galaxy mass buildup. Additionally, the use of adaptive optics (AO) on the ground \citep{schreiber14, Schreiber18, Jafariyazani19} and slitless spectroscopy from space, including observations with HST \citep{nelson12, nelson13, nelson16, Tacchella18, xinwang20, Simons21, Matharu22, Li_22} and JWST \citep{Matharu23, Li23, Nelson23, Matharu24, Liu24}, has provided a comprehensive picture of galactic mass assembly through the study of ionized gas emission lines. Combining these multi-wavelength observations, a physical picture emerges, explaining galaxy structural formation as a blend of extended, young star-forming disks and centrally concentrated, extremely dusty bulges.

However, most observations of molecular gas and dust utilizing ALMA have focused on very dusty and massive galaxies (e.g., log $\rm{M_*/M_\odot}$ $>$ 11.0) due to the prioritization of observation time. While these studies have illuminated the processes of galaxy formation in the high-redshift universe, our understanding of more typical main-sequence galaxies with moderate masses remains incomplete in the ALMA era. Main-sequence galaxies, which constitute the majority of galaxies throughout cosmic history, form stars in a steady, regulated manner with high (40-70\%) duty cycles of star formation rather than in sporadic bursts \citep[e.g.,][]{Brinchmann04, Noeske07}. This steady star formation likely arises from a balance between gas inflows, outflows, and star formation processes \citep[e.g.,][]{Daddi10, Tacconi10, Genzel2010, Tacconi13}. Understanding the internal life cycles of these galaxies is crucial because they dominate the star formation budget of the universe and provide key insights into typical galaxy evolution processes. Slitless spectroscopy from space and ground-based AO have helped decipher star formation in main-sequence galaxies by providing spatially resolved emission line maps. Yet, these studies are primarily conducted using \ha\ at $z>1$ due to wavelength constraints, which poses challenges. Although \ha\ is one of the best indicators of star formation, dust extinction can significantly attenuate its flux. This attenuation is challenging to assess because its effects are degenerate with those of stellar age and metallicity \citep[e.g.,][]{Schreiber09, Sobral12, Kashino13, Koyama19, Matharu23}.

Longer wavelength IR observations enabled by JWST have made significant strides in addressing these issues by providing high-resolution images from 0.6 to 28 $\mu$m. Wide-field slitless spectroscopy, now extended up to $\sim$ 5 $\mu$m, is a powerful tool that effectively functions as a redshift machine, providing unbiased identification of strong emitters within its field of view \citep[e.g.,][]{Sun22, Sun23, Matthee23, Matthee24, Helton24, Meyer24, Morishita24_a2744, Naidu24}. Additionally, its high spatial resolution opens a new window for studying spatially resolved line properties with unprecedented detail \citep[e.g.,][]{Nelson23, Matharu24, Liu24}. 
For the first time, we are able to spatially resolve \pa\ emission lines up to $z=1.7$. \pa\ traces the ionized gas and star formation activity with less sensitivity to dust attenuation due to its longer wavelength, thus serve as a robust tracer of star formation \citep{Kennicutt12, Reddy23}. Diverse star-formation patterns revealed by WFSS-based \pa\ spectra have been observed in massive galaxies, yet the physical mechanisms shaping these distinct star formation processes are not well understood \citep{Liu24}. Understanding such diversity requires studying molecular gas and gas dynamics. 

However, high spectral resolution of WFSS can lead to degeneracies between velocity and spatial information along the spectral axis, resulting in morphological distortions that complicate direct comparisons with continuum images or molecular gas maps. The forward modeling enabled by \texttt{Grizli} \citep{brammer22_grizli} can help decouple these degeneracies by combining paired direct imaging with grism exposures, though it relies on certain assumptions about the original emission line profile. To overcome this challenge, we present the first case study combining archival ALMA CO(2-1) observations with JWST NIRCam WFSS \pa\ emission line data for ASPECS-LP.3mm.06, a main-sequence spiral galaxy at $z=1.095$. This galaxy was spectroscopically confirmed with MUSE \citep{Bacon17} and detected in CO(2-1) through ALMA band 3 observations \citep{Jorge2019}. We explore the galaxy's molecular gas dynamics, spatial extent, ionized gas distribution, and SFE within the galaxy in a spatially resolved manner. Leveraging the kinematic information provided by ALMA, we implement two methods to break the degeneracies in the WFSS 2D spectra, allowing for a direct comparison with ALMA observations.

Our analysis provides insights into the potential star formation mode and the evolutionary stage of this galaxy. Although we analyze only one galaxy in this paper, which hampers us to draw any general conclusion, this work demonstrates the effectiveness of our new methods to directly compare ALMA and JWST WFSS data for spatially resolved studies.

Throughout the paper, we adopt the AB magnitude system \citep{oke83,Fukugita96}, cosmological parameters of $\Omega_m=0.3$, $\Omega_\Lambda=0.7$, $H_0=70$\,km\,s$^{-1}\,{\rm Mpc}^{-1}$, and the \citet{Chabrier03} initial mass function (IMF).

\begin{figure*}[ht]
    \centering
    \includegraphics[width=0.9\textwidth]{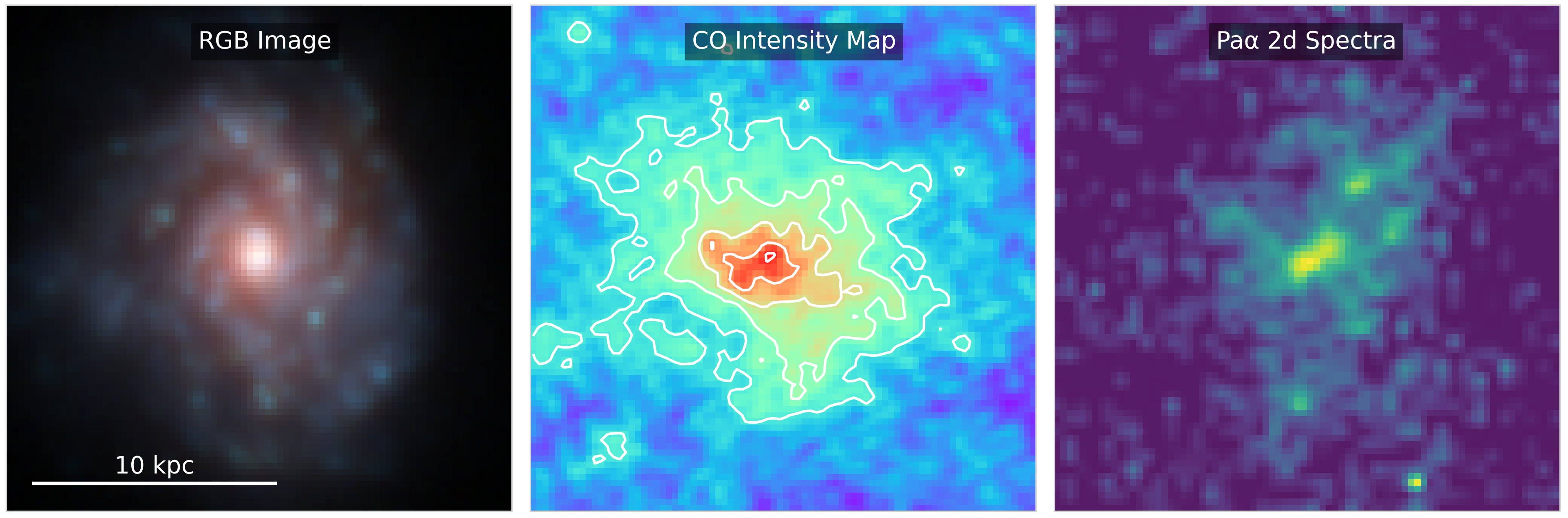}  
    \caption{Cutout stamps for ASPECS-LP.3mm.06, each measuring $2.\!''5 \times 2.\!''5$. 
    Left: JWST NIRCam RGB image generated using F115W, F150W, and F444W filters. 
    Middle: ALMA CO moment 0 map with integrated CO flux contours overlaid, starting at $4\sigma$ and increasing in steps of $2\sigma$. 
    Right: JWST WFSS \pa\ 2d spectra.}
    \label{fig:maps}
\end{figure*}

\section{Observations} \label{sec:data}

ASPECS-LP.3mm.06 is an ideal target for investigating the evolution and star formation processes shaping the main sequence at $z\sim1$, thanks to the availability of high-resolution data across multiple wavelengths. This galaxy has been covered by HST ACS/WFC3, JWST NIRCam imaging, NIRCam WFSS, and ALMA Band 3 spectroscopy. The combination of these datasets offers a unique opportunity to study the galaxy’s structure and properties with exceptional details, from the Near-UV (NUV) to millimeter wavelengths. The multi-wavelength data also allows us to resolve the galaxy’s morphology and physical processes with high precision, as illustrated in Fig. \ref{fig:maps}. Below, we outline the observational data used in this work:

\subsection{NIRCam+HST Photometry}\label{sec:nircam catalog}
The photometry is based on a catalog generated by \citet{Morishita24}, who utilized the methodology outlined in \citet{morishita23} and implemented through {\tt borgpipe} \citep{morishita21}. This catalog includes available HST ACS/WFC3 and JWST NIRCam data, spanning wavelengths from 0.4 to 4.5 $\mu$m. The JWST data were obtained by accessing the fully processed images and spectroscopic catalogs provided by the JADES team (\citealt{Rieke23}), which include observations from nine NIRCam filters. In this analysis, we employ aperture fluxes from the catalog for spectral energy distribution (SED) fitting, which have been corrected to represent the total fluxes of the galaxies.

\subsection{HST and JWST Grism}\label{sec:clear}

We utilize NIRCam WFSS F444W data from the First Reionization Epoch Spectroscopic Complete Survey (FRESCO; GO-1895; PI: P. Oesch; \citealt{Oesch23}) to study the \pa\ emission line. The FRESCO spectra achieve a resolution of R $\sim$ 1600 and covers a wavelength range from 3.8 to 5.0 $\mu$m, which enables us to study multiple hydrogen recombination lines across a wide range of cosmic time with a spatially resolved view. Following the methodology outlined in \citet{Sun23}\footnote{\href{https://github.com/fengwusun/nircam\_grism/}{https://github.com/fengwusun/nircam\_grism/}}, we reduce the WFSS data using a combination of the official JWST pipeline and several customized steps. Starting with stage-1 products (\texttt{rate.fits}) from the MAST archive, we assign the world coordinate system (WCS) and apply flat fielding using the latest data from the JWST Calibration Reference Data System (CRDS). Background subtraction involves two steps: first, subtracting a median background created for each module and pupil, and second, using \texttt{SExtractor} \citep{bertin1996} to refine the subtraction. We extract the emission line by removing the continuum, modeled via a median filter with a smoothing kernel \citep{Kashino23_grism}. For a detailed description of our reduction process, we refer the reader to \citet{Liu24}.

To estimate the dust extinction, we utilize the \ha\ flux taken with HST G141 grism from the catalogs complied by \citet{simons23_clear} as part of the CANDELS Lyman Alpha Emission At Reionization survey (CLEAR; PI: Casey Papovich). The \ha\ + \nii\ complex is blended in the HST G141 grism. To correct for \nii\ contamination, we derive the metallicity based on MUSE observations \citep{Bacon17, Bacon23}, using the [Ne$\,${\footnotesize III}] and [O$\,${\footnotesize II}] line ratios following the calibration of \citet{Maiolino08}. With a measured [Ne$\,${\footnotesize III}]/[O$\,${\footnotesize II}] ratio of (3.22 $\pm$ 0.02) $\times 10^{-2}$, we determine a metallicity of $12 + \log(\text{O/H}) = 8.90$. We then also adopt the \citet{Maiolino08} calibration, resulting in a \nii/\ha\ ratio of 0.29. We apply this ratio to remove the \nii\ contribution. Both the \pa\ flux that we measured and the \ha\ flux presented in the \citet{simons23_clear} catalog were obtained via optimal extraction \citep{Horne86}. We use \pa/\ha\ line ratio to determine dust extinction, which will be detailed in Sec. \ref{sec:physical properties}.

\subsection{ALMA observations}\label{sec:alma}
Our target was detected with CO(2-1) emission through The ALMA Spectroscopic Survey in the Hubble Ultra Deep Field (ASPECS; \citealt{Walter16_aspecs}). ASPECS is a 3D survey of gas and dust in distant galaxies, aiming to provide an unbiased census of molecular gas and dust continuum emission in galaxies at $z > 0.5$. The ALMA Band 3 and Band 6 observations were obtained during Cycle 2 as part of projects 2013.1.00146.S (PI: F. Walter) and 2013.1.00718.S (PI: M. Aravena). The ASPECS CO observations revealed that our target, ASPECS-LP.3mm.06, exhibits rich CO emission and an extended structure. Building upon this foundation, follow-up observations were conducted with ALMA Band 3 during Cycle 6 as part of project 2018.1.01521.S (PI: A. Hygate). This project involved deeper and higher angular resolution CO(2-1) observations using both extended and compact array configurations.

\begin{figure}[ht]
    \centering
    \includegraphics[width=0.46\textwidth]{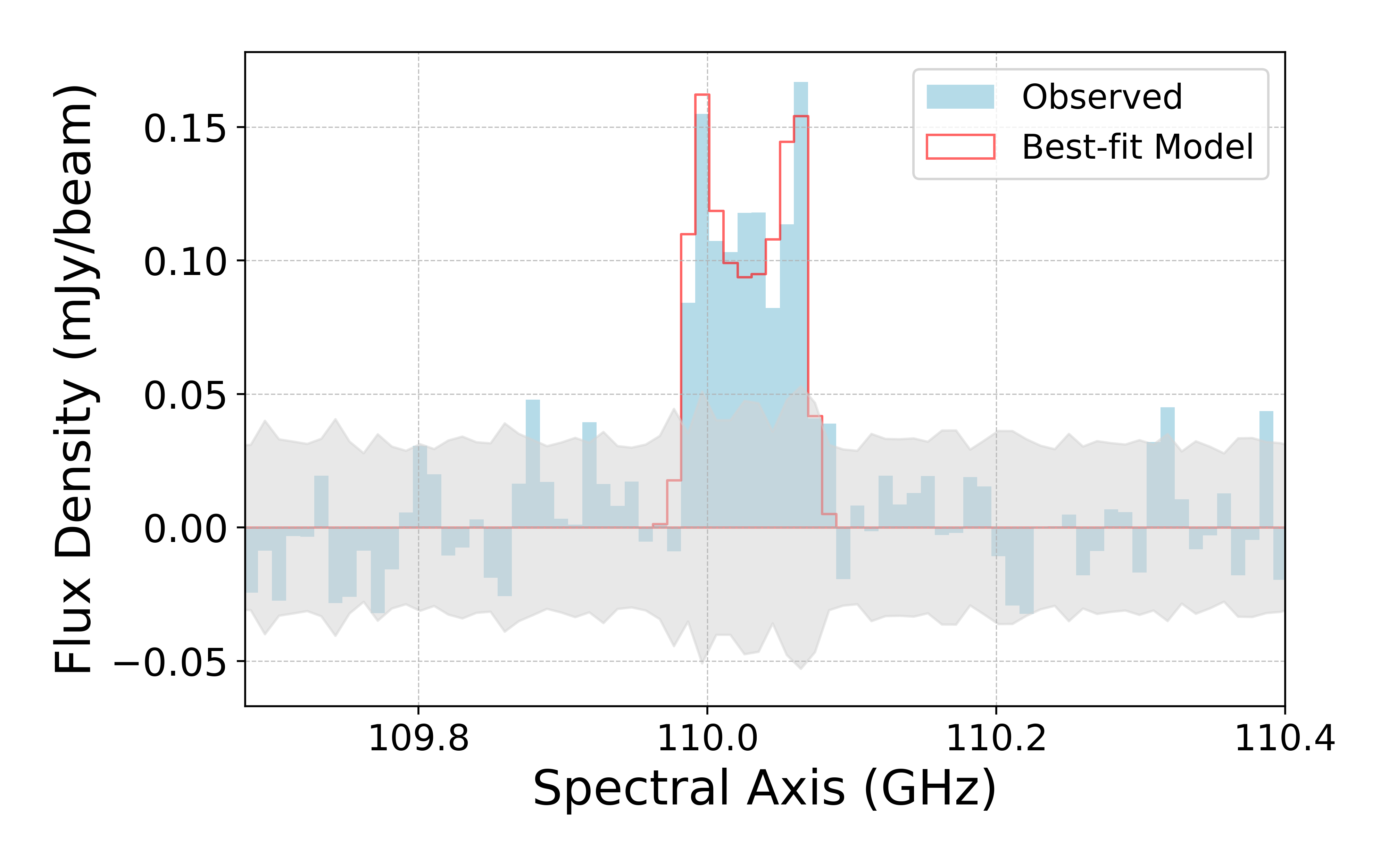}  
    \caption{ALMA CO(2-1) spectra of ASPECS-LP.3mm.06. The blue histogram represents the rebinned observed spectra with a channel width of 26.6 km/s, while the red line shows the best-fit model derived from our kinematic analysis (See Sec.~\ref{sec:kinematic model}; Fig.~\ref{fig:rotation}). The gray shaded region represents the 1$\sigma$ error for each bin.}
    \label{fig:co emission}
\end{figure}

\begin{figure}[ht]
    \centering
    \includegraphics[width=0.46\textwidth]{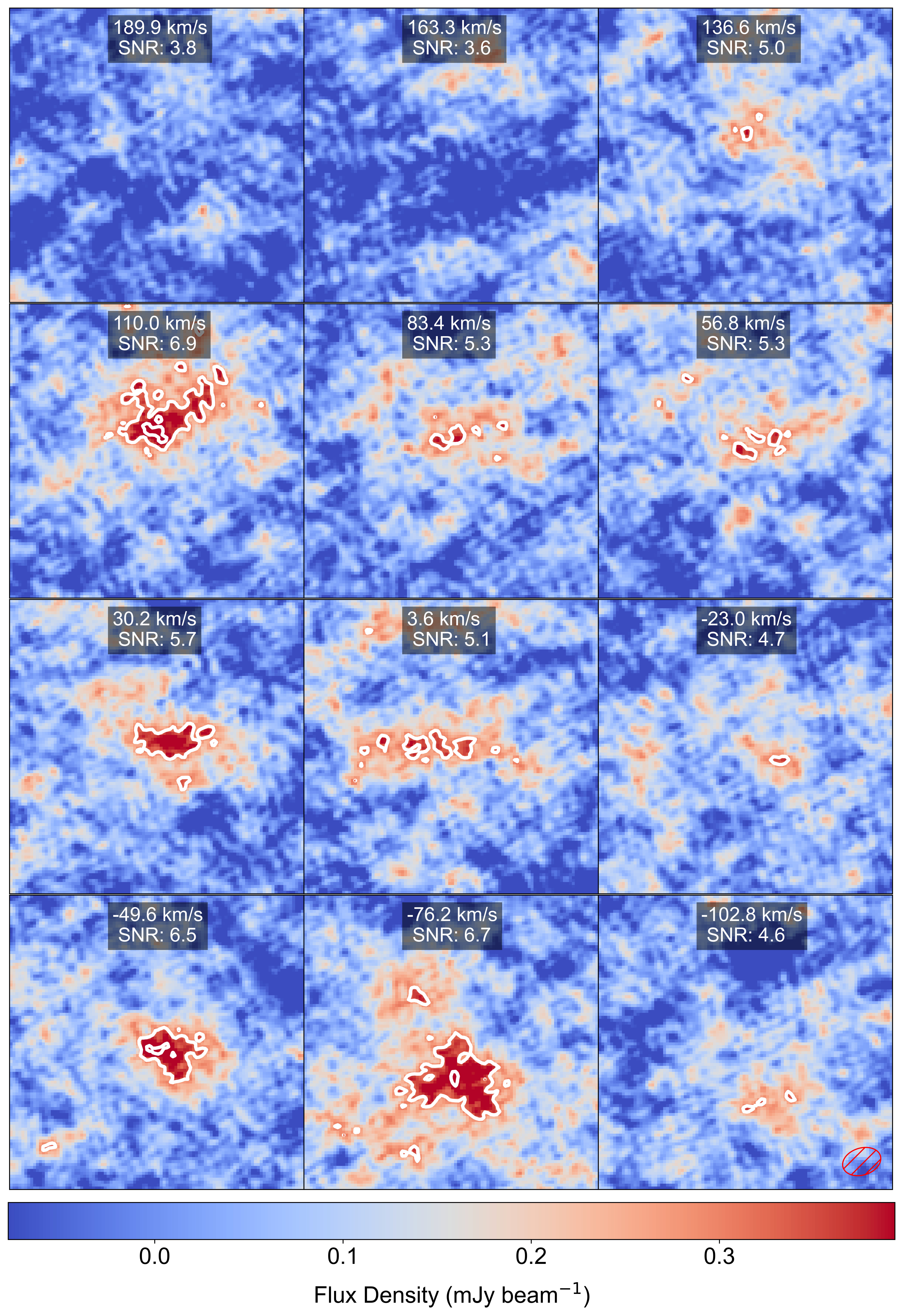}  
    \caption{Channel maps of the CO(2-1) emission from ASPECS-LP.3mm.06. The white contours indicate the CO(2-1) emission starting at $4\sigma$ and increasing in increments of $2\sigma$, where $\sigma$ represents the RMS noise level. In each panel, we note the velocity offset from the systemic velocity of the source, as well as the peak SNR for that specific channel. The red dashed ellipse in the bottom right panel represents the beam size.}
    \label{fig:channel}
\end{figure}

In this work, we primarily focus on project 2018.1.01521.S. We obtain the calibrated measurement sets from the East Asian ALMA Regional Center. We combine the measurement sets and conduct the data reduction using the Common Astronomy Software Applications (\texttt{CASA}; \citealt{McMullin07, CASA2022}), following standard calibration procedures. The original spectra, with a spectral resolution of 2.66 kms$^{-1}$, are rebinned to a channel width of 26.6 km s$^{-1}$. The extracted spectra and rebinned channels maps are shown in Fig. \ref{fig:co emission} and Fig. \ref{fig:channel}, respectively. The final processed data cube has a beam size of $0.\!''32 \times 0.\!''22$. Additionally, we create a separate cube using only the compact array configuration to calculate the total flux, with a beam size of $0.\!''81 \times 0.\!''66$, which will be revisited in Sec. \ref{sec:gas mass}.

\section{ANALYSIS AND RESULTS}

\subsection{Stellar Mass, Dust Extinction and Star Formation Rates}
\label{sec:physical properties}
We derive physical properties of our sample, including stellar mass, star formation rates (SFRs), and dust extinction, based on the HST and JWST observations described above. Our methodology involves the following steps:

\begin{enumerate}
    \item Stellar Mass Estimation:
    We estimate the stellar mass using the SED fitting tool \texttt{CIGALE} (Code Investigating GALaxy Emission; \citealt{2019A&A...622A.103B}) with avilable photometric observations from HST and JWST (Sec. \ref{sec:nircam catalog}). \texttt{CIGALE} fits the SED of galaxies to model their stellar populations and estimate their mass.

   \item Dust Extinction Correction:
    We estimate the dust extinction at \pa\ (A$_\mathrm{Pa\alpha}$) by using the observed flux ratio of Pa$\alpha$ (from the FRESCO survey) to H$\alpha$ (from the CLEAR survey) and comparing it to the theoretical ratio expected under Case B recombination conditions. By applying the \citet{Calzetti2000} reddening curve ($R_{\rm{V}} = 4.05$), we derive the dust extinction and use it to calculate the intrinsic \pa\ flux, which is then used to determine the SFR.

    \item Star Formation Rate Calculation: The SFR is estimated by first converting the dust-corrected \pa\ emission line flux to dust-corrected \ha\ line flux, assuming the Case B recombination scenario, which predicts an intrinsic intensity ratio of \ha/\pa\ = 9.15. This approach assumes an electron temperature of ${\rm T_e = 10^{4}}$ K and an electron density of ${\rm n_e = 10^{2}}$ ${\rm cm^{-3}}$, as determined by photoionization models using CLOUDY version 17.02 \citep{Ferland17}. The SFR is then estimated from the \ha\ luminosity following the SFR-$L_{\mathrm{H\alpha}}$ relation given by \citet{kennicutt98_2} and is scaled to align with the \citet{Chabrier03} IMF by dividing by a factor of 1.53 \citep{Driver13}.
    
\end{enumerate}


Our calculations yield a massive galaxy with M$_{*} = 3.7\times10^{10}\, M_\odot$ and a star formation rate of SFR$_\mathrm{Pa\alpha} = 28.0 \pm 1.0\, M_\odot\, \mathrm{yr}^{-1}$ after correcting for dust extinction, with a modest dust extinction at \pa\ of A$_\mathrm{Pa\alpha} = 0.13 \pm 0.03$ mag. As shown by \citet{Reddy23}, the \citet{Calzetti2000} extinction law remains applicable even at \pa\ wavelengths. Our measurement of A$_\mathrm{Pa\alpha} = 0.13 \pm 0.03$ mag corresponds to A$_\mathrm{V} = 0.90 \pm 0.21$ mag and is marginally consistent with the SED-based A$_\mathrm{V} = 0.8$ reported by \citet{Aravena20}, further validating the use of the \citet{Calzetti2000} law for longer-wavelength lines such as \pa. For a detailed explanation of our measurements, we refer readers to \citet{Liu24}.

\subsection{Kinematic Modeling} \label{sec:kinematic model}
\begin{figure*}[ht]
    \centering
    \includegraphics[width=0.9\textwidth]{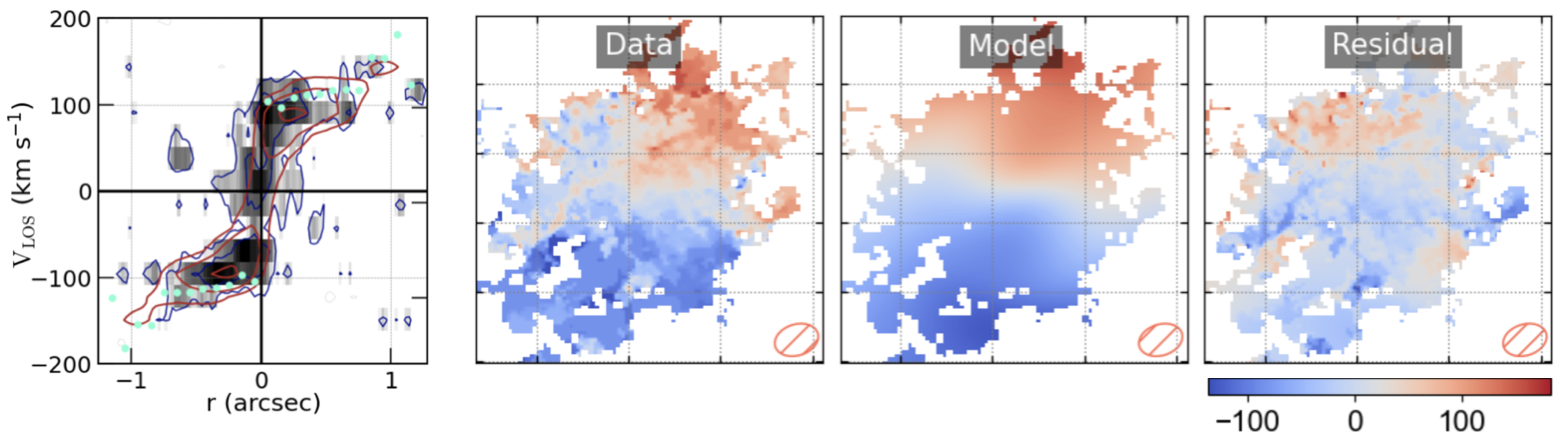}  
    \caption{Kinematics of ASPECS-LP.3mm.06. The left panel shows the model and observed velocities extracted along the major axis on the empirical position-velocity (P-V) diagram. The moment maps, with dimensions \(2.\!''5 \times 2.\!''5\), are presented in three columns: the first column displays the observed data, the second column shows the corresponding model, and the third column presents the residuals between the data and the model. The red dashed ellipse corresponds to the beam size.}
    \label{fig:rotation}
\end{figure*}

We further study the kinematic properties of the galaxy using the $\mathrm{^{3D}}$\texttt{BAROLO} software (\texttt{BBarolo}, \citealt{Teodoro15}), which fits the emission line data cube by employing tilted ring models. This method approximates the galaxy's disk as a series of concentric rings, each with its own kinematic and geometric parameters. We follow the fitting process as described in \cite{Pope23} and start with using \texttt{SEARCH} algorithm to mask pixels below 3$\sigma$ detection. We fit both the kinematic and geometric parameters using a ring width of $r_{\mathrm{ring}} = 0.\!''1$. The ring width is chosen to be the FWHM of the minor axis of the restoring beam divided by 2.2, which is comparable to other ring width ratios reported in the literature (2.2, \citealt{Shao17}; 2.5, \citealt{Jones21}). We assume a thin disk and fix the scale height of the rings to Z$^o$ =  0.\!''01, We note that assuming a thicker disk (2.5$\times$, 5$\times$, or 10$\times$) does not affect the final fitting results. The initial parameter guesses, including the central position, rotation velocity, and velocity dispersion, are based on the moment maps. In the first fit, we set both geometric and kinematic parameters as free variables. We then take the best-fit parameters from this initial fit and use them in a subsequent fit, where only the maximum velocity and dispersion are allowed to vary. The best-fit velocity map is shown in Fig. \ref{fig:rotation}.

We derive the kinematic parameters based on the fitting results. We estimate the $V_{\text{max}}$ by taking the average rotational velocity of all rings, resulting in $V_{\text{max}} = 253.7 \pm 31.0$ km$^{-1}$. The velocity dispersion $\sigma_{0}$ is calculated by taking the average dispersion over all rings with nonzero dispersion, considering their respective uncertainties, yielding $\sigma_{0} = 27.6 \pm 8.3$ km$^{-1}$. Given that $V_{\text{max}}/\sigma_{0} = 9.2 \pm 3.0$, ASPECS-LP.3mm.06 is consistent with dynamically cold, rotating disk galaxy.



\subsection{Dust Continuum, Gas Mass and Star Formation Efficiency}
\label{sec:gas mass}
We use the spectral windows where the emission line does not fall to estimate the continuum level. Our target is undetected in the continuum with a sensitivity of $\sigma = 3\,\mu$Jy/beam. Based on the ASPECS observations, ASPECS-LP.3mm.06 was undetected in the 3 mm continuum and only detected in the deep 1.2 mm continuum \citep{Jorge2019, Aravena20, Jorge2020}. \citet{Aravena20} also estimated the physical properties of ASPECS galaxies using \texttt{magphys} \citep{Cunha08}, a SED fitting routine that uses a prescription to balance the energy output at various wavelengths. This method produces estimates of the the dust extinction $A_\mathrm{V}$ and SFR, which are compiled in Table~\ref{tab:physical_properties}.

We estimate the molecular gas mass (M$_{mol}$) using the CO(2-1) emission line flux. As mentioned in Sec. \ref{sec:alma}, for the emission line flux calculation, we use only the compact array (low resolution) configuration. This decision is due to the incorrect default flux scaling of the CLEAN residual map, known as the ``JvM effect'' \citep{Jorsater95}, which may overestimate the flux in the CLEANed map when combining two configurations. The problem arises because the dirty beam resulting from the combination of two arrays is not as Gaussian as one assumes when using the clean beam. In our case, the compact array configuration already achieves a good SNR of 12.1 for calculating the total flux. Therefore, we use only this single configuration for the total flux calculation to avoid the issues associated with the JvM effect.

The velocity-integrated CO(2-1) emission line flux of our sample galaxy is $S_{\text{int}} =\,(0.54 \pm 0.04)$\,Jy\,km\,s$^{-1}$, consistent with the values reported in the ASPECS observations within the uncertainties \citep{Jorge2019, Aravena19}, which measured an integrated flux density of $S_{\text{int}} =\,(0.48 \pm 0.06)$\,Jy\,km\,s$^{-1}$ for the same target. To convert CO(2-1) to CO(1-0), we use the excitation ladder estimated by \citet{Daddi15}, where $\mathrm{r_{21}=0.76}$. To estimate the total molecular gas mass from the CO(1-0) luminosity, we apply a CO-to-H${2}$ conversion factor ($\alpha_{\mathrm{CO}}$), which accounts for its dependence on gas-phase metallicity, as $\alpha_{\mathrm{CO}}$ is known to decrease with increasing metallicity \citep[e.g.,][]{Wilson95, Leroy11, Genzel15, Tacconi18}. Using the metallicity we derived for our galaxy from the [Ne$\,${\footnotesize III}]/[O$\,${\footnotesize II}] ratio (see Sec. \ref{sec:clear}), and applying the metallicity-dependent calibration from \citet{amorin16}, we obtain $\alpha_{\mathrm{CO}} = 2.45$. As a result, the total molecular gas mass of our galaxy is $M_\mathrm{{gas}} = (2.9\,\pm\,0.2)\times10^{10}M_{\odot}$. This measurement is consistent with that reported by \citet{Aravena20} using multiple independent methods (SED-based, Rayleigh-Jeans dust-based, CO-based).



We then estimate the SFE for the entire galaxy using the dust-corrected \pa\ based SFRs (see Sec. \ref{sec:physical properties}) and CO(2-1) based molecular gas mass using:

\begin{equation}
\text{SFE} = \frac{\text{SFR$_\mathrm{Pa\alpha}$}}{M_{\text{gas}}}
\end{equation}

The calculated star formation efficiency of ASPECS-LP.3mm.06 is $\text{SFE} = (9.7\,\pm\,0.7)\times10^{-10}\,yr^{-1}$.



Based on the analysis in Sec.~\ref{sec:physical properties}, \ref{sec:kinematic model}, \ref{sec:gas mass}, ASPECS-LP.3mm.06 appears to be a massive, moderately dusty main sequence galaxy with a stable rotating gas disk at $z_{\mathrm{CO}}=1.0951$/ $z_{\mathrm{Pa\alpha}}=1.0955$. The physical properties of this galaxy, as estimated from this work and the ASPECS observations \citep{Aravena19, Boogaard19}, are compiled in Table~\ref{tab:physical_properties}.

    

\begin{table}[h]
    \centering
    \caption{Physical Properties of ASPECS-LP.3mm.06}
    \resizebox{0.43\textwidth}{!}{ 
    \begin{threeparttable}
    \begin{tabular}{lll}
        \hline
        {Physical Properties} & {Value} & {Unit} \\
        \hline
        M$_{*}$ & $3.7\times10^{10}$  & $M_\odot$\\
        M$_{mol}$ & $(2.9\pm0.2)\times10^{10}$ & $M_\odot$\\ 
        $S_{\text{int}}$ & 0.54 $\pm$ 0.04 & Jy\,km s$^{-1}$\\
        f$_\mathrm{mol}$ \tnote{a}& 0.44 $\pm$ 0.02 & \\
        Observed SFR$_\mathrm{Pa\alpha}$ & 24.8 $\pm$ 0.8 & $M_\odot$ yr$^{-1}$\\
        Intrinsic SFR$_\mathrm{Pa\alpha}$ & 28.0 $\pm$ 1.0 & $M_\odot$ yr$^{-1}$ \\
        SED based SFR \tnote{b}& 34.0 & $M_\odot$ yr$^{-1}$\\
        A$_\mathrm{Pa\alpha}$ & 0.13  $\pm$ 0.03 & mag \\
        A$_\mathrm{V}$  \tnote{b}& 0.8 & mag\\
        V$_\mathrm{max}$ & 253.7 $\pm$ 31.0 & km s$^{-1}$\\
        $\sigma_{0}$ &  27.6 $\pm$ 8.3 & km s$^{-1} $\\
        \hline 
    \end{tabular}
    \begin{tablenotes}
    \item[a] f$_\mathrm{mol}$ = M$_{mol}$ / ( M$_{*}$ +  M$_{mol}$)
    \item[b] The values were derived by \citet{Aravena19, Aravena20, Boogaard19} based on ASPECS observations.

    \end{tablenotes}
    \end{threeparttable}
    }
    \label{tab:physical_properties}
\end{table}

\subsection{Size Measurements}
\label{sec:size}
Leveraging the extended structure and gas-rich disk of ASPECS-LP.3mm.06, along with high-resolution archival ALMA and JWST observations, we are able to resolve its critical properties, including stellar light, ionized gas, and molecular gas. 


We first compare the flux distribution of F090W, F444W, and the original CO(2-1) moment 0 map. The F090W image corresponds to the rest-frame $\sim$ 0.4 $\mu$m, falling within the NUV range. The F444W image captures the rest-frame K-band ($\sim$ 2 $\mu$m), which is an accurate probe of the galaxy's stellar mass \citep[e.g.,][]{Kauffmann98, 2001ApJ...550..212B, Drory04}. 

Given that ASPECS-LP.3mm.06 is an extended source with a complex spiral arm structure that may not be adequately captured by a simple S$\acute{\mathrm{e}}$rsic model \citep{1963BAAA....6...41S}, we employ a PSF deconvolution technique to measure its size in the NIRCam images, following the method of \citet{Szomoru10}, utilizing \texttt{WebbPSF} \citep{2012SPIE.8442E..3DP, 2014SPIE.9143E..3XP} and \texttt{Galfit} \citep{peng2002, peng2010}. First, we generate the PSF for the corresponding filter using \texttt{WebbPSF}. We then perform S$\acute{\mathrm{e}}$rsic $R^{1/n}$ fitting with \texttt{Galfit} to obtain the best-fit parameters by minimizing the residuals between the PSF-convolved model and the observed data. The deconvolved model is the intrinsic (PSF-free) S$\acute{\mathrm{e}}$rsic profile generated from these best-fit parameters. By subtracting the PSF-convolved best-fit model from the observed data, we obtain residuals that capture deviations from the S$\acute{\mathrm{e}}$rsic model. We then add these residuals back to the deconvolved S$\acute{\mathrm{e}}$rsic model to reconstruct the intrinsic flux distribution. This combined profile is used to derive the intrinsic flux profile for size measurement. Despite residual PSF effects, this technique effectively reconstructs the true flux distribution, even for galaxies that deviate from a S$\acute{\mathrm{e}}$rsic profile \citep{Szomoru10}.

Finally, we measure the half-light radius ($r_{1/2}$) from the azimuthally averaged radial profile of the deconvolved image. To estimate the uncertainty in $r_{1/2}$, we perform Monte Carlo simulations by generating 100 model galaxies. For each simulation, we randomly sample the S$\acute{\mathrm{e}}$rsic parameters from normal distributions centered on the best-fit values obtained from \texttt{Galfit}, with standard deviations equal to their uncertainties. We create deconvolved S$\acute{\mathrm{e}}$rsic models using these sampled parameters and add the corresponding residuals from the original fit to capture deviations from the S$\acute{\mathrm{e}}$rsic model. We then measure the half-light radius for each of these modeled galaxies. The uncertainty is derived from the standard deviation of the half-light radii obtained from the simulations. By doing so, our measurement is not affected by PSF effects or the limitations of using a S$\acute{\mathrm{e}}$rsic $R^{1/n}$ model to represent the complete structure.

For the CO(2–1) map, it is possible to obtain an effective radius using the same method as for JWST by deconvolving the image with the synthesized beam and modeling the deconvolved image with a two-dimensional profile. However, the image reconstruction for interferometric data depends heavily on the uv-coverage of the visibilities and the CLEANing process. These factors can introduce uncertainties in galaxy size measurements. Therefore, we measure the size directly from the visibility data using the \texttt{CASA} package \texttt{uvmodelfit} assuming Gaussian profile. Previous studies have shown that comparing image-based size measurements with visibility-based size measurements is valid and does not introduce significant systematic uncertainties \citep[e.g.,][]{Hodge16, Chen20,tadaki20, Ikeda22}. The effective radii calculated for F090W, F444W, and CO are listed in Table~\ref{tab:effect_radii}.



\begin{table}[h]
    \centering
    \caption{Half-light Radius of ASPECS-LP.3mm.06}
    \begin{threeparttable}
    \begin{tabular}{llll}
        \hline
        {F090W} & {F444W} & {CO(2-1)} & {Unit} \\
        \hline
        5.29 $\pm$ 0.27 & 4.15 $\pm$ 0.18 & 5.72 $\pm$ 0.28 & {kpc}\\
        \hline 
    \end{tabular}
    \end{threeparttable}
    \label{tab:effect_radii}
\end{table}

The images and the half-light radius ($r_{1/2}$) both indicate that our target exhibits extended emission, particularly in the F090W and CO map. This observation suggests active ongoing star formation in the galaxy disk, which is fueled by sufficient cold gas. The F444W image, with its relatively low half-light radius ($r_{1/2}$), could be indicative of the presence of a bulge emitting central stellar light, resulting in a smaller radius. The relatively large effective radius in the F090W can also be attributed to central flux attenuation, which, if present, may flatten the flux profile.

\subsection{Spatially Resolved Gas, Star Formation and SFE}

We further study the flux distribution of the ionized gas probe, i.e., the \pa\ 2D spectra, and the molecular gas tracer, i.e., the CO moment 0 map. However, the NIRCam grism achieves a medium spectral resolution of $R \sim 1600$ at 3.95 $\mu$m, corresponding to a velocity resolution of $\sigma \sim 76.3$ km s$^{-1}$. This resolution may result in morphological distortions in the emission line maps along the spectral axis, making it difficult to directly compare the \pa\ emission with the integrated CO moment 0 map. 

To break the velocity and morphology degeneracy in WFSS, several approaches have been proposed. In a recent work, \citet{Nelson23} utilized an additional medium-band filter to extract emission line maps. By comparing the medium-band based emission line maps with the WFSS-extracted 2D spectra, they were able to obtain the velocity gradient. However, this approach significantly narrows the applicable redshift range due to the limited coverage of medium-band filters and is primarily applicable to high EW galaxies with robust detection in the medium-band only. Dynamical forward modeling is another potential solution for obtaining separate kinematic and morphological information. \citet{Li23} studied a $z = 8.34$ galaxy using WFSS and F480M, where they convolved the F480M-derived 2D brightness profile with the rotation velocity and velocity dispersion fields, and used MCMC sampling to derive the best-fit kinematic parameters. However, there are still uncertainties regarding the use of medium-band images as emission line maps for this purpose, as the actual emission line map can have a different flux distribution compared to the medium-band captured continuum \citep[e.g.,][]{nelson12, nelson16, Vulcani16, Matharu21}.

To overcome these difficulties and more accurately investigate the spatial distribution of molecular gas, ionized gas, and SFE, we employ two approaches that enable direct comparison of WFSS-based 2D spectra with CO line map. Each has its own advantages and disadvantages. The first approach involves shifting the ALMA channel map to manually match the spectral extent of the \pa\ 2D spectra, enabling a direct comparison. The second approach involves reconstructing the \pa\ emission line map using the velocity information from CO observations, under the assumption that the molecular gas and ionized gas follow the same kinematic features. In both methods, we use the \pa\ 2D spectra without correcting for spatially resolved dust extinction.

\subsubsection{Velocity-shifted CO map}
\label{Velocity-shifted CO map}

\begin{figure*}[ht]
    \centering
    \includegraphics[width=0.95\textwidth]{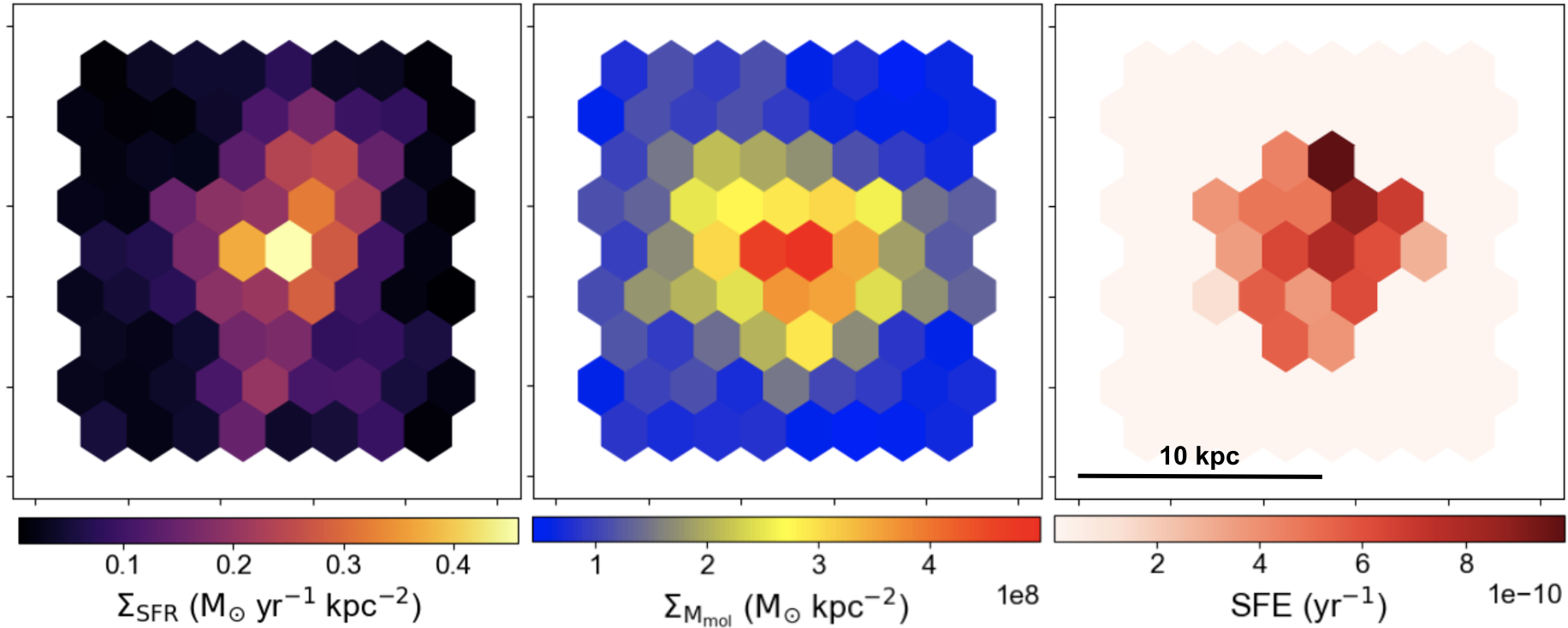}  
    \caption{Hexagonal binned cutout stamps showing \pa\ 2d spectra (left),  velocity-shifted CO moment 0 map (middle) and SFE map (right) for ASPECS-LP.3mm.06, each measuring 
$2.\!''5 \times 2.\!''5$.}
    \label{fig:sfe}
\end{figure*}
In the first method, where we shift the ALMA channel map to match the \pa\ 2D spectra, we rerun \texttt{CASA} and adjust the relevant parameters in the \texttt{tclean} task to bin the channel maps into 77.1 km s$^{-1}$ intervals, This is the closest value to the spectral resolution of WFSS, which is 76.3 km s$^{-1}$, given the original ALMA data's spectral resolution of 2.66 km s$^{-1}$. Consequently, we obtain three channel maps at velocities of -74.3 km s$^{-1}$, 2.8 km s$^{-1}$, and 79.9 km s$^{-1}$, with the zero velocity position defined as the center of the CO emission line. We then shift the first channel map along the spectral axis by $0.\!''063$ in the wavelength increasing direction and the third channel map by $0.\!''063$ in the wavelength decreasing direction. The shift of $0.\!''063$ corresponds to one pixel size of the \pa\ 2D spectra. The central channel map is kept unchanged. These shifted channel maps are then stacked to mimic the \pa\ dispersed spectra, and we refer to the resulting map as the ``velocity-shifted moment 0 map'' in the following discussion. This is to differentiate it from the original moment 0 map, which does not contain any kinematic information.

Since the moment 0 map is shifted to match the \pa\ spectra, both images now contain spectral and spatial information aligned along the dispersion direction in the same way. This alignment allows us to directly compare their flux distributions without concern for morphological distortions along the spectral axis. Both the velocity-shifted moment 0 map and the Pa$\alpha$ emission line map are extended, showing ongoing star formation in the galaxy's central bulge and several spiral arms. To further understand how SFE varies across different sites within the galaxy and to mitigate the variance in single pixels, we employ hexagonal binning. This involves mapping the original pixel data from rectangular grids to hexagonal grids. Such a binning strategy is suitable for our purpose as it provides a more stable and representative measure of SFE, allowing us to smooth out noise and uncover underlying patterns more effectively.

We calculate the SFE in each hexagon as the ratio of \pa\ flux to CO flux. The velocity-shifted moment 0 map, \pa\ 2D spectra, and corresponding SFE map are shown in Fig. \ref{fig:sfe}. To avoid potential issues in regions with low SNR, such as where a small SFR is divided by a small molecular gas mass, we constrain the region used for calculating SFE to those with a 4$\sigma$ CO detection and exclude low CO SNR regions when displaying SFE. As a result, the SFE map appears smooth across both the disk and bulge regions, suggesting relatively stable growth in both the bulge and disk, with no evidence of a central starburst. However, we note that the depth of the \pa\ 2D spectra is shallower compared to the CO integrated map, and the \pa\ spectra exhibit more clumps rather than a smooth disk, similar to what has been observed in other \pa\ emission line maps \citep{Liu24, Naidu24}. The existence of these small clumps may lead to an underestimation of the SFE, particularly in the outer disk regions. Although the irregular dust distribution within the galaxy, such as heavier extinction in the bulge, could also contribute to an underestimation of SFE in that area, it is important to note that this galaxy is not particularly dusty. Our SFR tracer, \pa, observed at a longer wavelength, has an extinction value of A$_\mathrm{Pa\alpha}$ = 0.13, resulting in only a 11\% flux difference before and after extinction correction. Therefore, we conclude that using the \pa\ map, even without correcting for extinction, provides a reliable approximation and does not significantly impact our results.


\begin{figure*}[ht]
    \centering
    \includegraphics[width=0.92\textwidth]{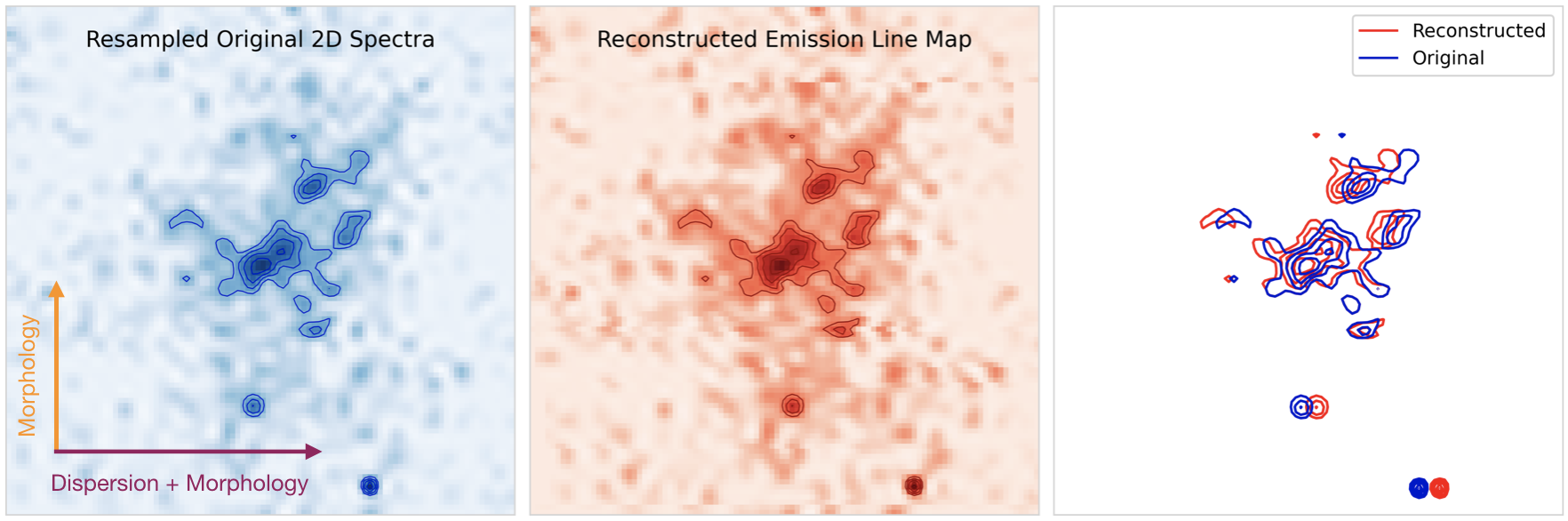}  
     \caption{Resampled original 2D spectra (left) and reconstructed emission line map (middle), with associated contours starting at $3\sigma$ and increasing in steps of $1\sigma$. The right panel compares the contours representing the reconstructed emission line map (red) and the distorted original spectra (blue). The cutout size of each map corresponds to $2.\!''5 \times 2.\!''5$.}
    \label{fig:emline}
\end{figure*}

\begin{figure*}[ht]
    \centering
    \includegraphics[width=0.95\textwidth]{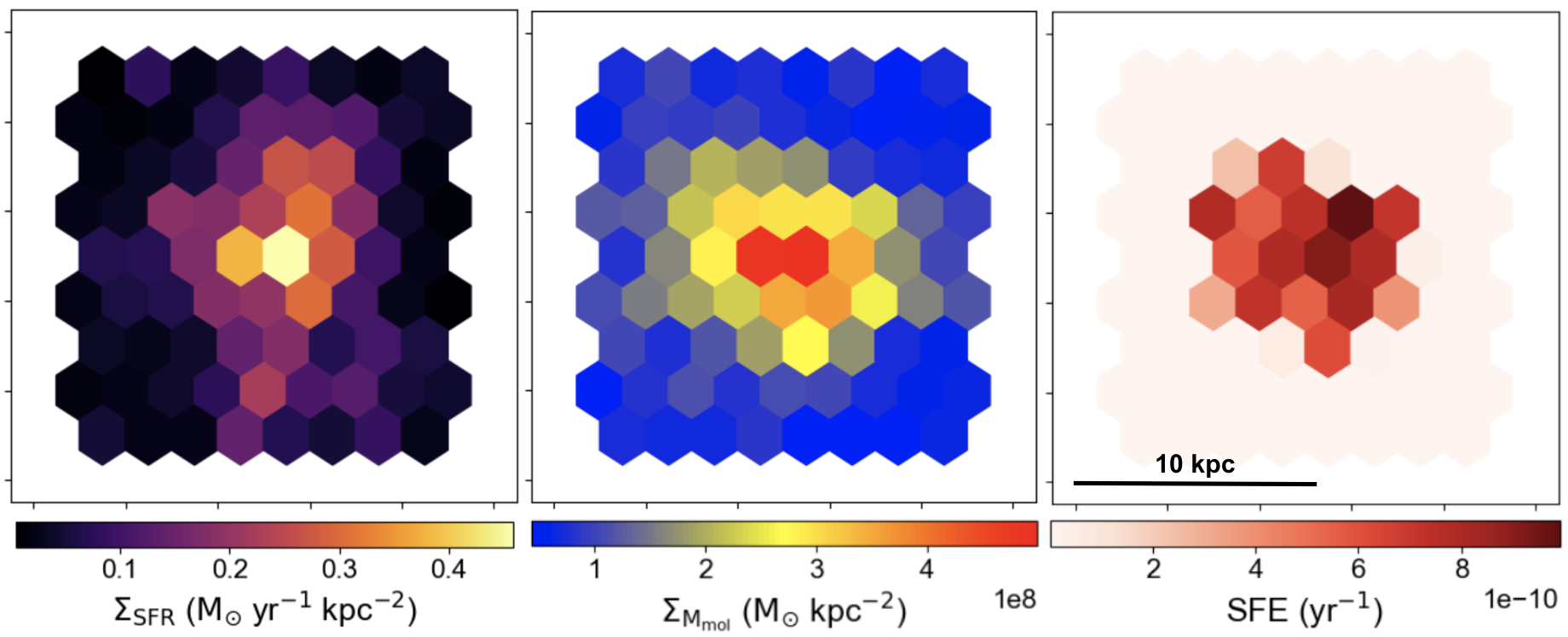}  
    \caption{Hexagonal binned cutout stamps showing \pa\ reconstructed emission line map (left),  original CO moment 0 map (middle) and SFE map (right) for ASPECS-LP.3mm.06, each measuring 
$2.\!''5 \times 2.\!''5$.}
    \label{fig:sfe2}
\end{figure*}

\begin{figure}[ht]
    \centering
    \includegraphics[width=0.48\textwidth]{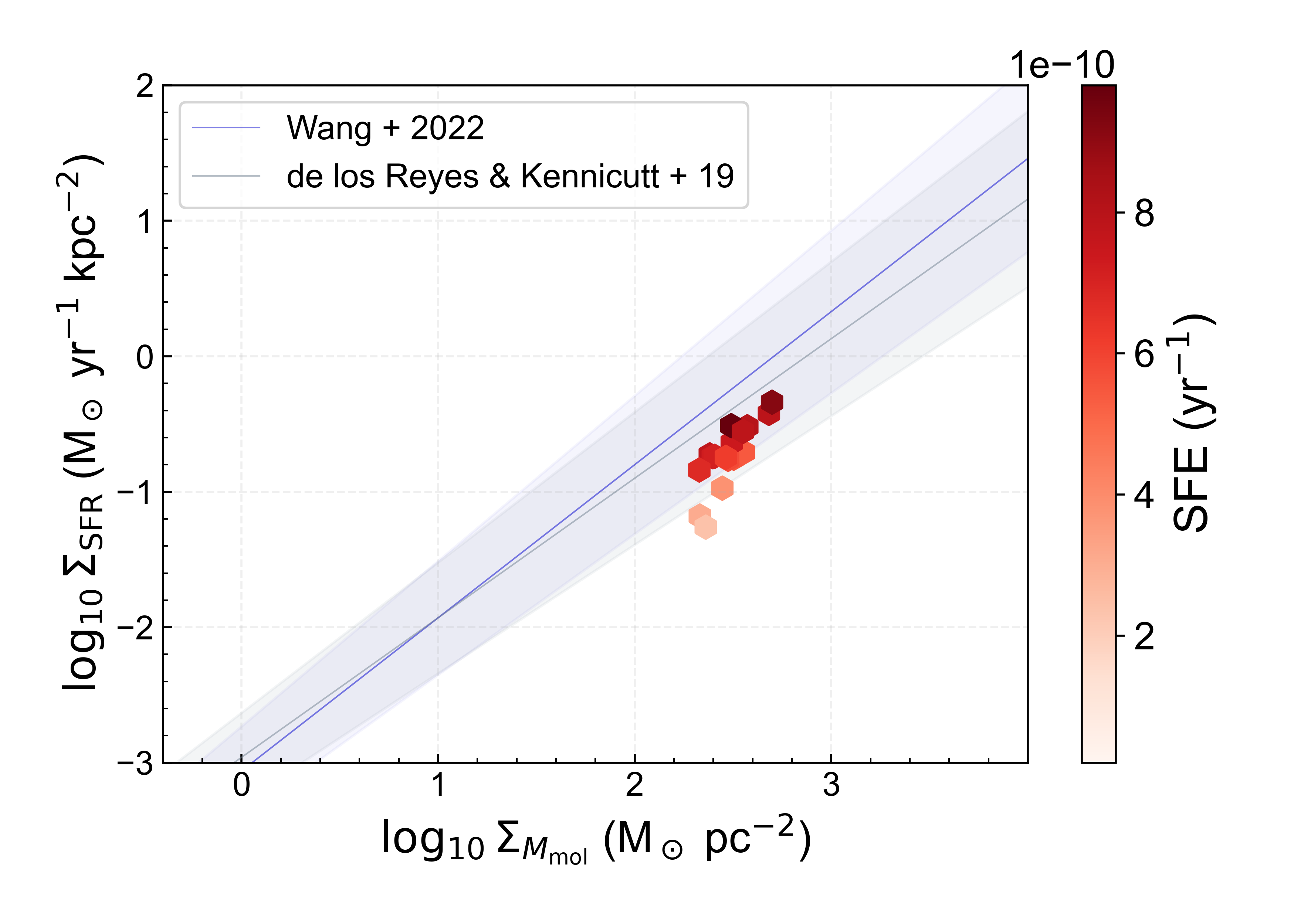}  
    \caption{Kennicutt-Schmidt relation of hexagonal binned regions in ASPECS-LP.3mm.06. The color of each hexagon represents their SFE, as displayed in Fig. \ref{fig:sfe2}. Only the regions with CO detection above 4$\sigma$ are shown. The blue line is the best-fit KS relation obtained by \citet{Wang22} for MS SFGs at $z = 0.4 - 3.6$. The gray line is the best-fit KS relation obtained for local galaxies by \citet{delosReyes19}. The shaded regions represent the uncertainties associated with each respective slope.}
    \label{fig:ks}
\end{figure}

\subsubsection{\texorpdfstring{Reconstructed \pa\ Emission Line Map}{Reconstructed pa Emission Line Map}}
In the second method, we reconstruct the \pa\ emission line map using the 2D spectra and the CO velocity. Since the velocity dispersion of our target galaxy is smaller than the velocity gradient per pixel ($\sim$ 76.3 km s$^{-1}$), we consider only the effect of the line-of-sight rotational velocity derived from the best-fit moment 1 model, as detailed in Sec. \ref{sec:kinematic model}.

The core concept of this method is to model the galaxy as a collection of distinct segments, each with its own velocity, leading to varying shifts in the 2D spectra. Our goal is to correct these shifts by leveraging both the 2D spectra (\pa\ 2d spectra) and the velocity information provided by the CO moment 1 map. While it is in principle possible to perform this reconstruction on a pixel-by-pixel basis, the differing velocities among pixels within the same row could result in overlapping values in the final 2D spectra. To avoid this complication, we opt to treat each row of the 2D spectra as a separate segment. By calculating the average velocity for the corresponding regions in the moment 1 map, we accurately determine the extent to which each row is shifted along the dispersion direction and apply the necessary corrections.

To implement this, we first resample the \pa\ 2D spectra to match the pixel scale of the CO map (from $0.\!''063$ to $0.\!''025$ per pixel). This resampling reduces the velocity gradient to $\sim$ 30.3 km s$^{-1}$ per pixel in the resampled image, ensuring that the 2D spectra and the CO map are pixel-to-pixel aligned. Next, we calculate the shift in the 2D spectra for each row by dividing the average rotation velocity by the velocity gradient per pixel and rounding the result to the nearest integer. Rows with a positive rotation velocity, indicating a shift towards redder wavelengths (increasing wavelength direction), are shifted back accordingly. Similarly, rows with negative rotation velocity are shifted in the opposite direction. This process allows us to reverse the velocity-induced shifts and reconstruct the intrinsic emission line map.

The \pa\ 2D spectra before and after reconstruction are presented in Fig. \ref{fig:emline}. We perform the same SFE calculation as described in Sec. \ref{Velocity-shifted CO map}, with the results displayed in Fig. \ref{fig:sfe2}. While the SFE obtained through the two methods shows slight differences, there are intrinsic distinctions between the two SFE maps. Specifically, the method using the velocity-shifted CO map and the original 2D spectra derives an SFE map that includes the galaxy's velocity information. In contrast, the method using the original CO map and the reconstructed \pa\ map derives an intrinsic SFE map without velocity information. This can lead to the \pa\ star-forming clumps falling within different hexagonal bins in the resulting SFE maps, contributing to the observed differences between the two methods. Moreover, uncertainties introduced during the shifting process can also contribute to discrepancies in the SFE maps. Despite these differences, our conclusion remains consistent: the SFE does not change significantly across both the disk and bulge regions, regardless of the method employed.

With the reconstructed emission line map and the original CO moment 0 map, we can also investigate the Kennicutt-Schmidt (KS) law of our target galaxy in a spatially resolved manner. As shown in Fig. \ref{fig:ks}, the hexagonally binned regions of ASPECS-LP.3mm.06 are plotted alongside the best-fit KS laws from \citet{Wang22} and \citet{delosReyes19}. Across different regions, our target galaxy's areas are positioned on or slightly below the KS relation, indicating homogeneous star formation conditions with typical or slightly lower gas-to-star conversion rates.

\section{Discussions}
\label{sec:discussion}

\subsection{Star Formation in Gas-Rich Disk Galaxies}
\label{sec:discussion1}

\begin{figure}[ht]
    \centering
    \includegraphics[width=0.48\textwidth]{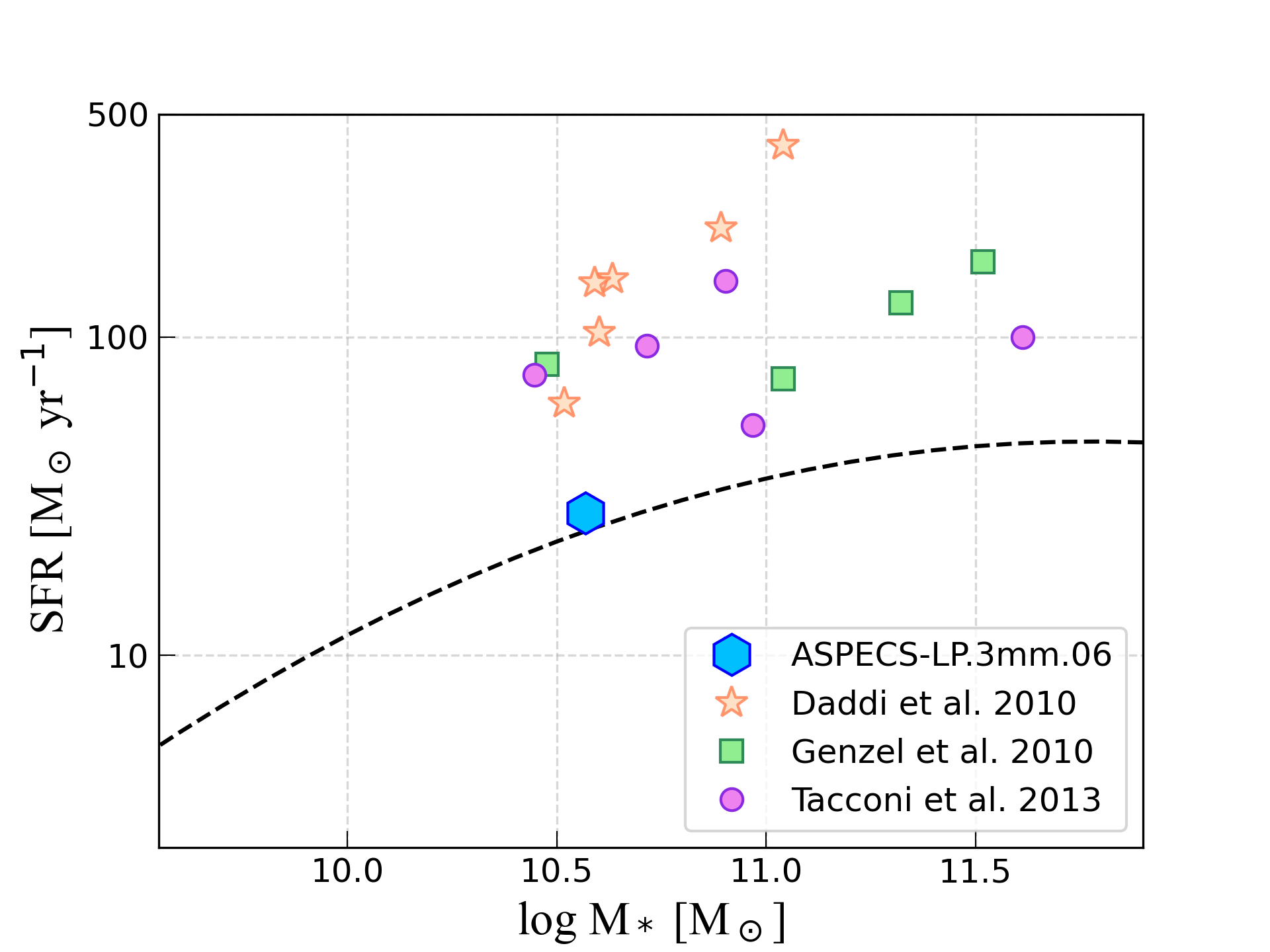}  
    \caption{Position of ASPECS-LP.3mm.06 on the star-forming main sequence, compared with literature values for galaxies at $1 < z < 2$ with molecular gas fractions $f_\mathrm{mol} > 0.4$ \citep{Daddi10, Genzel2010, Tacconi13}. The dashed black line corresponds to the main sequence relation at $z=1.1$ \citep{Popesso23}.}
    \label{fig:ms}
\end{figure}

The mode of star formation that feeds the growth of massive disk galaxies is still under debate. Thanks to the rich archival data set, we are able to constrain the star formation mode in ASPECS-LP.3mm.06. As a representative star-forming galaxy at this cosmic epoch, demonstrated by its position on the main sequence of star-forming galaxies (Fig. \ref{fig:ms}), detailed study of ASPECS-LP.3mm.06 can provide valuable insights into the physical processes governing similar galaxies at this redshift and stellar mass. In the following, we explore several scenarios that feed disk galaxy growth and their potential relation to ASPECS-LP.3mm.06.



Disk instabilities are closely tied to star formation because they create dense regions that can collapse under their own gravity to form new stars. In gas-rich galaxies, the high surface density exacerbates these instabilities, leading to the gravitational fragmentation of gas-rich, thick turbulent disks into large clumps and other dense structures. This increased turbulence further promotes star formation, resulting in rapid star formation and the formation of compact bulges \citep[e.g.,][]{Toomre64, Wang94, Noguchi99, Dekel2009}. Observational evidence supports this scenario, as many sub-millimeter galaxies are reported to have centrally concentrated star formation and gaseous clumps \citep[e.g.,][]{Tacconi08, Tadaki18}, which are often interpreted to involve gas-rich major mergers. However, it has been found that star formation in some sub-millimeter galaxies is rather complex and highly degenerate, making it difficult to attribute it to a single scenario. For instance, \citet{Hodge12} studied GN-20, a sub-millimeter galaxy at $z \sim 4$, and found that it shows multiple clumps along with a stable rotating disk, indicating features of both a gas-rich merger and fueling by some process other than a major merger. In addition to the gas-rich merger, different characteristics of gas accretion to discs also produce different star forming activities in galaxies. One hypothesis assumes the existence of cold-mode accretion (CMA; \citealt{2005MNRAS.363....2K, Dekel09}), where gas remains cold and directly accretes onto galaxies through gas filaments without being heated up. CMA has been believed to be an crucial formation scenario for the origin of massive disk galaxies, it is particularly important in low-mass halos and at high redshifts, where cold gas can flow directly into the galaxy, fueling intense star formation. In contrast, in more massive halos, gas is shock-heated to the virial temperature before slowly accreting onto the galaxy, leading to reduced star formation activity. Therefore, CMA can explain both the rapid growth of star-forming galaxies at their early formation stage, as well as the observed quenching in more massive galaxies \citep{Noguchi23}. \citet{Daddi10} studied six normal, near-infrared selected (BzK) galaxies at $z \sim 1.5$ and reported extended CO reservoirs with spatial sizes consistent with what is measured from their stellar light, which disfavors violent mergers characterized by highly concentrated molecular gas distributions. Similar findings for intermediate redshift normal star-forming galaxies have been reported  \citep[e.g.,][]{Genzel06, Daddi08, Tacconi10}, further supporting the CMA model. The properties of ASPECS-LP.3mm.06 align with galaxies that are thought to be fueled by CMA. 

Fig. \ref{fig:ms} illustrates the position of ASPECS-LP.3mm.06 on the main sequence alongside galaxies from the literature that exhibit comparable molecular gas fractions (f$_\mathrm{mol} > 0.4$) and similar redshifts ($1 < z < 2$). It is important to note that, though the gas fraction of ASPECS-LP.3mm.06 is comparable to those reported in the literature \citep{Daddi10, Genzel2010, Tacconi13}, the SFR of ASPECS-LP.3mm.06 is slightly lower. \citet{Daddi10}'s galaxies in a similar mass range exhibit SFRs ranging from 62 to 400 $M_\odot \text{ yr}^{-1}$, while ASPECS-LP.3mm.06 has an \pa-based SFR of 28.0 $M_\odot \text{ yr}^{-1}$. This lower SFR may be attributed to the fact that the galaxies studied by \citet{Daddi10} are more turbulent, leading to the formation of clumpy star-forming structures. In contrast, ASPECS-LP.3mm.06 is more dynamically stable, exhibiting orderly rotational dynamics, lacking giant clumps, and having star formation distributed more evenly across smaller clumps throughout the galaxy.

This picture is reinforced by recent JWST observations, which have revealed that disk galaxies were already in place by $z>1.5$ \citep{Ferreira22,Robertson23, Lee24_disk}. Many of these galaxies are gas-rich and dynamically cold, with star formation occurring in undisturbed disks, suggesting that major mergers may not be the primary mechanism driving the formation of massive disk galaxies \citep[e.g.,][]{cheng23, Gillman24}.

\subsection{Moderately Low Dust Extinction}

Another interesting property of ASPECS-LP.3mm.06 is its modest dust reservoir. Gas-rich galaxies with active ongoing star-forming activities are generally expected to have bright FIR emission and strong dust extinction, as gas mass and dust mass are well-established to be correlated with each other at a given metallicity \citep{Issa90, Galametz11, Ruyer14}. MUSE observations using [Ne$\,${\footnotesize III}] and [O$\,${\footnotesize II}] reveal that ASPECS-LP.3mm.06 is a metal-rich galaxy, with $12 + \log(\text{O/H}) = 8.90$. Given the high gas fraction and metallicity, it would be expected for ASPECS-LP.3mm.06 to be a dustier galaxy. However, it was undetected in the 3 mm continuum and only detected in the very deep 1.2 mm continuum \citep{Jorge2019, Jorge2020}. Furthermore, both SED-based (A$_\mathrm{V}$=0.8 mag) and hydrogen recombination line-based extinction values (A$_\mathrm{Pa\alpha}=$0.13$\pm$0.03 mag) suggest mild dust content. A recent study by \citet{Hodge24} examined 13 infrared-luminous massive galaxies. One of their samples, ALESS 17.1 at $z \sim 1.5$, exhibits a bright dust continuum and a nearby bright spiral-like stellar structure with an offset of $\sim 0.\!''8$ at the same redshift. This structure is interpreted as a merger. In contrast, our target, ASPECS-LP.3mm.06, shows no significant signs of ongoing mergers or interactions, which may be responsible for its relatively low dust content. Despite being a major population in galaxy evolution, galaxies like ASPECS-LP.3mm.06 are still underrepresented in many previous ALMA surveys due to their relatively low dust content and moderate SFRs comparable to the MS.
\begin{figure}[ht]
    \centering
    \includegraphics[width=0.46\textwidth]{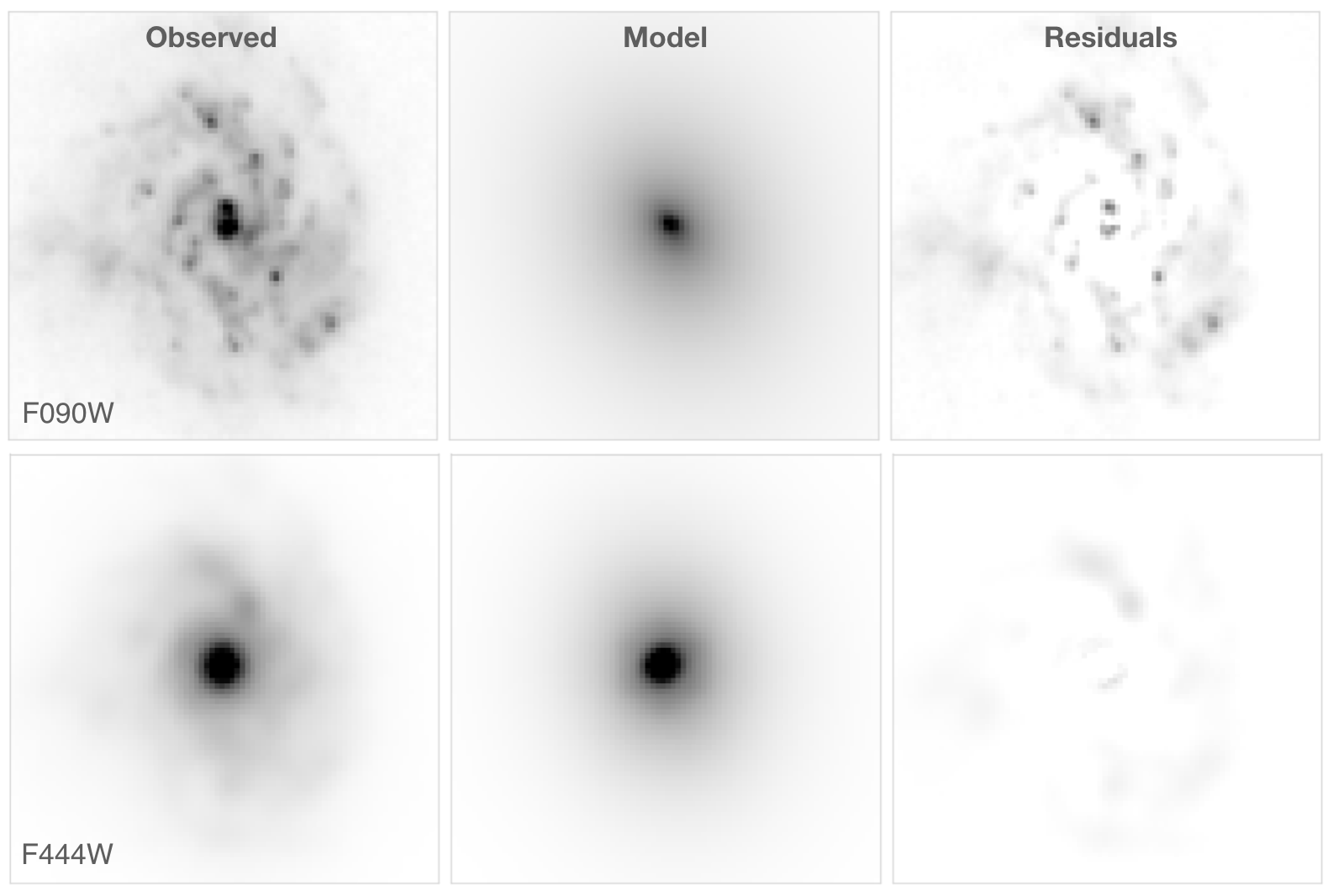}  
    \caption{Two-component S$\acute{\mathrm{e}}$rsic fitting results for F090W (top) and F444W (bottom). From left to right, each column represents the observed images, model fits, and residuals.}
    \label{fig:galfit}
\end{figure}

\begin{figure}[ht]
    \centering
    \includegraphics[width=0.46\textwidth]{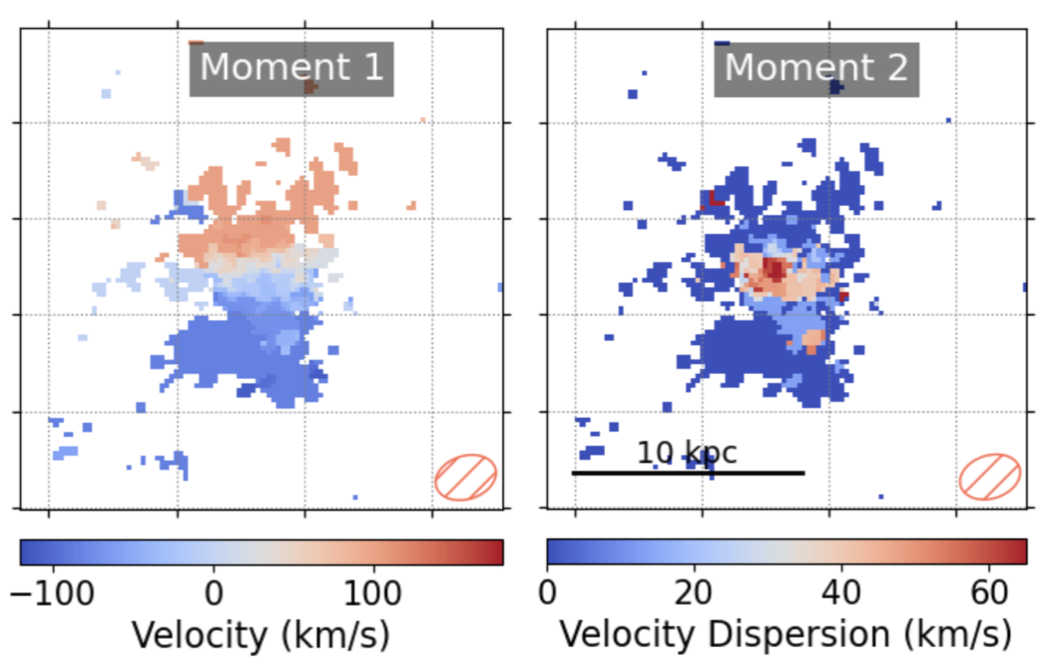}  
    \caption{Observed moment 1 (left) and moment 2 (right) maps.}
    \label{fig:velocity dispersion}
\end{figure}

Additionally, the low dust extinction in our target can be explained by the modest star formation in the bulge region. Many heavily dust-attenuated galaxies exhibit significant compact dust emission within the central $\sim$1 kpc, indicating ongoing strong starburst activities associated with a larger amount of dust \citep{Tadaki17, Hodge19, tadaki20, Chen20}. These observations suggest that the presence of compact dust emission and efficient star formation are linked to the formation of central dense stellar structures, i.e., galaxy bulges. The lack of significant dust emission in ASPECS-LP.3mm.06 might indicate that its bulge formation is not as active as in dusty starburst galaxies, resulting in less central dust concentration.


To quantify the bulge fraction, we conduct two-component S$\acute{\mathrm{e}}$rsic \citep{1963BAAA....6...41S} fitting using {\tt{galfit}} \citep{peng2002, peng2010}. As illustrated in Fig. \ref{fig:galfit}, the central light concentration is more prominent in the F444W filter compared to F090W. In contrast, F090W reveals a greater number of small star-forming clumps, emphasizing regions of active star formation distributed across the disk. Since F444W serves as an approximate tracer of the galaxy's stellar mass, we derive the bulge-to-total (B/T) ratio using the {\tt GALFIT} model based on the F444W light distribution. In our fitting, the S$\acute{\mathrm{e}}$rsic indices are fixed at $n=4$ for the bulge and $n=1$ for the disk components. The derived B/T ratio of 0.13$\pm$0.01 indicates that ASPECS-LP.3mm.06 is primarily disk-dominated but hosts a modest bulge component. This low B/T ratio suggests that the galaxy is less likely to experience morphological quenching \citep{Martig09, Bluck14}. Although the B/T ratio is not high enough to classify the galaxy as an early-type, the presence of a bulge remains a significant indicator of its evolutionary state. During bulge formation, the galaxy mass becomes more centrally concentrated, increasing its gravitational potential. A stronger gravitational potential forces stars and gas to move faster to remain in orbit, leading to higher velocity dispersion. This increased velocity dispersion can cause the gas to become dynamically heated or turbulent, making it difficult for the gas to collapse under its own gravity and form stars \citep{Krumholz05}. This self-regulating process eventually halts star formation in the bulge as the galaxy reaches a more mature stage. ASPECS-LP.3mm.06 exhibits a significantly higher velocity dispersion ($>60$ km s$^{-1}$) within the central 1 kpc, while the velocity dispersion in the disk is much lower ($<10$ km s$^{-1}$), as shown in Fig. \ref{fig:velocity dispersion}. As such, the elevated velocity dispersion in the bulge explains the limited efficiencies in SFE in the central region, despite the abundant gas, and also accounts for the lack of a dusty starburst that could rapidly build up the bulge. A recent study by \citet{liu23_smacs} utilized NIRCam and MIRI to examine the spatial distribution of PAH emission and stellar continuum in galaxies at $1 < z < 1.7$. They found that most of their galaxies exhibited more compact PAH emission compared to the stellar continuum, suggesting active central star formation and bulge growth. In contrast, the most massive galaxy in their sample ($M_* \sim 10^{10.9}\,M_\odot$) showed similar spatial extents for both stellar and dust components, which could indicate a transition in star formation activity or the onset of quenching in the central regions. Although ASPECS-LP.3mm.06 differs from these galaxies in terms of dust content, the elevated central velocity dispersion and reduced SFE in the bulge imply that similar self-regulating processes during bulge formation may be occurring, leading to suppression of central star formation and the development of a more quiescent bulge.


In contrast, centrally concentrated starburst activity associated with bulge formation is commonly observed in many submillimeter galaxies (SMGs), which are significantly more massive and have more prominent bulge components than ASPECS-LP.3mm.06. This difference can be explained by distinct star formation modes: SMGs often undergo merger-driven, efficient bulge formation leading to intense central starbursts, while in normal main-sequence galaxies like our target, bulge formation is driven by gas inflows and disk instabilities, which are less efficient and result in a more gradual bulge build-up. The suppression of star formation and the low dust content could potentially be explained by the presence of an AGN or supernova feedback, as strong radiation fields can lead to the removal of central gas and dust or the destruction of dust grains \citep{Genzel98, Fabian08, Priestley21}. However, based on the analysis by \citet{Boogaard19}, although ASPECS-LP.3mm.06 is detected in X-rays, it is not classified as an AGN according to \textit{Chandra} X-ray observations \citep{Luo17}. Therefore, the absence of central starburst activity in this galaxy might instead be related to turbulence in the bulge region, which could be heating the molecular gas and contributing to the observed X-ray emission, similar to the suppressed star formation associated with warm molecular gas reported by \citet{Lanz16}. A stable gas inflow in the past may have sustained star formation in a smoother, more regulated manner, rather than through a turbulent, merger-driven process. As the bulge gradually grew, self-regulation in the bulge region could have suppressed further star formation, eventually leading the galaxy toward quenching.

\section{Summary and Future Prospect}
\label{sec:summary}
This paper presents a joint analysis of a $z \sim 1.1$ spiral galaxy, ASPECS-LP.3mm.06, utilizing JWST NIRCam WFSS and ALMA observations. By combining these two datasets, we were able to study the molecular gas dynamics and spatially resolved star formation on kiloparsec scales within this gas-rich, main-sequence galaxy.
 ASPECS-LP.3mm.06 was robustly detected in both the \pa\ emission line and CO(2-1) emission line. The \pa\ emission reveals an extended and clumpy structure, while the CO(2-1) emission indicates an extended gas disk. It exhibits a gas-rich nature with M$_{mol}$ = $(2.9\pm0.2)\times10^{10}$ $M_\odot$ and a moderate SFR with SFR$_\mathrm{Pa\alpha}$ = 28.0 $\pm$ 1.0 $M_\odot$ yr$^{-1}$. The size measurements of the galaxy reveal an extended structure in F090W (5.29 $\pm$ 0.27 kpc), F444W (4.15 $\pm$ 0.18 kpc), and the CO integrated map (5.72 $\pm$ 0.28 kpc). The F444W band, corresponding to the stellar mass distribution, has a slightly smaller half-light radius, which may be explained by the presence of a bulge. We modeled the kinematics of ASPECS-LP.3mm.06 and found a rotating disk with a flat rotation curve and regular kinematic features, classifying it as a dynamically cold disk galaxy. We developed two methods to enable direct comparison of NIRCam WFSS and ALMA observations. Based on this, we calculated the spatially resolved SFE within the galaxy and found a smooth SFE across both the bulge and disk regions. Despite the high gas fraction of 0.44 and high metallicity, ASPECS-LP.3mm.06 is not very dusty, with A$_\mathrm{Pa\alpha}$ = 0.13 and A$_\mathrm{V}$ = 0.8, distinguishing it from common sub-millimeter dusty starburst galaxies. The relatively low dust content, combined with the presence of a modest bulge component (with a bulge-to-disk ratio of 0.13$\pm$0.01), suggests that star formation in this galaxy may be regulated by central gas turbulence. The relatively low SFR of ASPECS-LP.3mm.06, compared to galaxies with similar stellar mass and gas fraction in the literature, further supports this hypothesis. Our observations suggest that ASPECS-LP.3mm.06 is a prototypical disk galaxy at z$\sim$1, with no clear evidence of recent major merger activity. The galaxy's current state may reflect an evolutionary path characterized by gradual bulge growth and self-regulated star formation. This indicates that internal processes, rather than external interactions, play a significant role in their evolution toward quiescence.


The successful combination of ALMA and JWST NIRCam WFSS observations in this study demonstrates the powerful synergy of these facilities in probing galaxy evolution during cosmic noon. The high-resolution, spatially resolved spectroscopy provided by JWST WFSS, together with ALMA's sensitivity to cold molecular gas, enables detailed studies of the interplay between star formation and gas dynamics on kpc scales. Future surveys employing both ALMA and JWST WFSS will provide a more comprehensive view of main-sequence galaxy build-up during this critical epoch, enabling investigations into the mechanisms driving star formation efficiency, bulge growth, and the galaxies' transitions to quiescence.

We sincerely thank the anonymous referee for the insightful comments, which have greatly enhanced the quality of this manuscript. We acknowledge the teams of the JWST observation programs \#1180, \#1895, and the HST observation program \#14227 for their hard work in designing and planning these programs, and for generously making their data publicly available. ZL would like to thank Dr.\,Jorge Gonz{\'a}lez-L{\'o}pez for discussions regarding data reduction and Ryota Ikeda for discussions on size measurement. The data presented in this paper were retrieved from the Mikulski Archive for Space Telescopes (MAST) at the Space Telescope Science Institute, the specific observations analyzed can be accessed via \dataset[DOI: 10.17909/z0sb-mk09]{http://dx.doi.org/10.17909/z0sb-mk09}. 
This paper makes use of the following ALMA data: ADS/JAO.ALMA\#2018.1.01521.S ALMA is a partnership of ESO (representing its member states), NSF (USA) and NINS (Japan), together with NRC (Canada), MOST and ASIAA (Taiwan), and KASI (Republic of Korea), in cooperation with the Republic of Chile. The Joint ALMA Observatory is operated by ESO, AUI/NRAO and NAOJ. Data analysis was carried out on the Multi-wavelength Data Analysis System operated by the Astronomy Data Center (ADC), National Astronomical Observatory of Japan. ZL is supported by KAKENHI Grant No. 24KJ0394 from the Japan Society for the Promotion of Science (JSPS) and by the ALMA Japan Research Grant of the NAOJ ALMA Project, NAOJ-ALMA-342. This work is also supported by JSPS KAKENHI Grants No. 24H00002 (Specially Promoted Research by T. Kodama et al.) and No. 22K21349 (International Leading Research by S. Miyazaki et al.).



\software{\texttt{Astropy} \citep{Astropy13,Astropy18},
          \texttt{CIGALE} \citep{2019A&A...622A.103B},
          \texttt{CASA} \citep{McMullin07, CASA2022},
          \texttt{Galfit} \citep{peng2002, peng2010}},
          \texttt{BBarolo} \citep{Teodoro15})

\bibliography{sample631}{}
\bibliographystyle{aasjournal}



\end{document}